\documentclass[sigconf]{acmart}
\AtBeginDocument{%
  }

\setcopyright{acmlicensed}
\copyrightyear{2026}
\acmYear{2026}
\setcopyright{cc}
\setcctype{by}
\acmConference[WSDM '26]{Proceedings of the Nineteenth ACM International Conference on Web Search and Data Mining}{February 22--26, 2026}{Boise, ID, USA}
\acmBooktitle{Proceedings of the Nineteenth ACM International Conference on Web Search and Data Mining (WSDM '26), February 22--26, 2026, Boise, ID, USA}
\acmPrice{}
\acmDOI{10.1145/3773966.3779406}
\acmISBN{979-8-4007-2292-9/2026/02}

\usepackage[ruled,vlined]{algorithm2e}
\usepackage{multirow}
\usepackage{hyperref}

\begin{document}

\title{A Real-Time System to Populate FRA Form 57 from News}

\author{Chansong Lim}
\affiliation{%
  \institution{University of California, Riverside}
  \city{Riverside}
  \state{California}
  \country{USA}
}
\email{clim090@ucr.edu}

\author{Haz Sameen Shahgir}
\affiliation{%
  \institution{University of California, Riverside}
  \city{Riverside}
  \state{California}
  \country{USA}
}
\email{hshah057@ucr.edu}

\author{Yue Dong}
\affiliation{%
  \institution{University of California, Riverside}
  \city{Riverside}
  \state{California}
  \country{USA}
}
\email{yued@ucr.edu}

\author{Jia Chen}
\affiliation{%
 \institution{University of California, Riverside}
  \city{Riverside}
  \state{California}
  \country{USA}
}
\email{jiac@ucr.edu}

\author{Evangelos E. Papalexakis}
\affiliation{%
  \institution{University of California, Riverside}
  \city{Riverside}
  \state{California}
  \country{USA}
}
\email{epapalex@cs.ucr.edu}

\renewcommand{\shortauthors}{Chansong Lim, Haz Sameen Shahgir, Yue Dong, Jia Chen, and Evangelos E. Papalexakis}

\begin{abstract}
  Local railway committees need timely situational awareness after highway–rail grade crossing incidents, yet official Federal Railroad Administration (FRA) investigations can take days to weeks. We present a demo system that populates \emph{Highway–Rail Grade Crossing Incident Data (Form 57)} from news in real time. Our approach addresses two core challenges: the form is visually irregular and semantically dense, and news is noisy. To solve these problems, we design a pipeline that first converts Form 57 into a JSON schema using a vision language model with sample aggregation, and then performs grouped question answering following the intent of the form layout to reduce ambiguity. In addition, we build an evaluation dataset by aligning scraped news articles with official FRA records and annotating retrievable information. We then assess our system against various alternatives in terms of information retrieval \emph{accuracy} and \emph{coverage}.
\end{abstract}




\maketitle

\section{Introduction}\label{intro}

Timely situational awareness is critical for local railway workers when a railway accident occurs. Although the Federal Railroad Administration (FRA) conducts investigation into such accidents, the official report can take from several days to multiple weeks to complete. During this latency, news articles are often the only publicly available sources of incident details. Our goal in this work is to monitor train accident news in real time and automatically extract information that can populate the \href{https://railroads.dot.gov/sites/fra.dot.gov/files/2019-09/FRA%20F%206180.57.pdf}{\emph{Highway–Rail Grade Crossing Accident/Incident Report (Form 57)}}, thereby providing local decision makers with real-time, referenced information.

\begin{figure}
    \centerline{\includegraphics[width=1\linewidth]{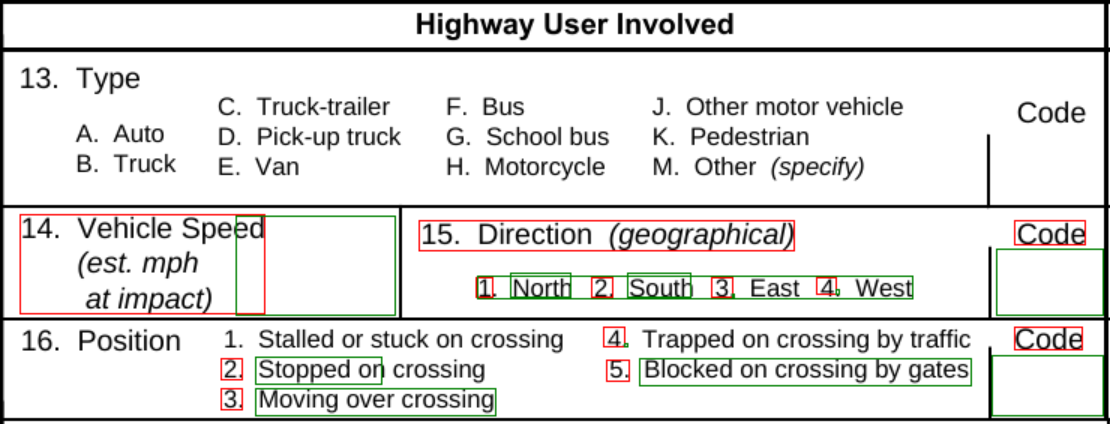}}
    \caption{Sample output from a widely used form parser (AWS Textract), which is one of the best working parsers. Red boxes denote keys and green boxes denote values that are extracted. As shown, the parser fails to extract complete key–value pairs under complicated conditions.}
    \label{fig: existing form parsers}
\end{figure}

\begin{figure*}[!htbp]
    \centerline{\includegraphics[width=1\linewidth]{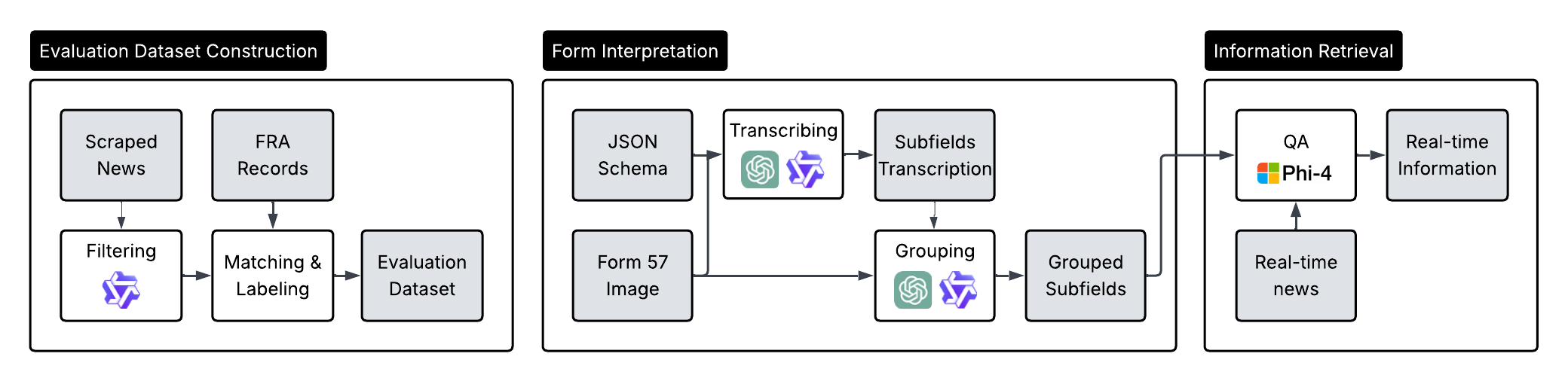}}
    \caption{The entire pipeline consists of form interpretaion, i.e. KIE, and IR. For evaluation of this system, we constructed a news dataset linked to FRA official report records and annotated manually.}
    \label{fig: pipeline}
\end{figure*}

Populating this form presents several challenges, particularly in understanding the form itself and retrieving relevant information. First, unlike other grid-structured forms, Form 57 is highly unstructured and individual cells often contain heterogeneous elements -- field labels, unordered choice lists, units of measure, checkboxes, and so on -- which cause traditional form parsers to misinterpret signals, as illustrated by the \href{https://docs.aws.amazon.com/textract/latest/dg/what-is.html}{AWS Textract} example in Fig~\ref{fig: existing form parsers}. Second, many fields are semantically ambiguous in isolation. For example, we can identify that “Direction” (Entry \#15) refers to a highway vehicle rather than rail equipment, only when surrounding fields provide contextual clues. Without such disambiguation, incorrect information may be extracted during information retrieval (IR).

To address these issues, we propose a pipeline consisting of: (1) Form57-to-JSON conversion with sample-aggregation using a vision–language model (VLM), and (2) grouped question answering (QA) that incorporates both the layout and semantics of the form. We categorize the fields because semantically related fields are typically positioned close to one another -- a design convention common to most document forms.

For evaluation, we collected news articles related to highway–railway accidents in California since 2000. Because such accidents seem to be relatively underreported -- we could just scrape 154 among 3,962 official records -- we collected as many articles as possible across counties and cities. After preprocessing, we aligned the articles with the corresponding official accident records and annotated which fields are retrievable from the news articles. We then evaluated our proposed system on this curated dataset.

Our contributions are summarized as follows:
\begin{itemize}
    \item We construct a railway accident news dataset linked with official records, despite the scarcity and noisiness of available news articles.
    \item We introduce develop a task-specific VLM formulation to parse a highly unstructured, information-rich form into structured JSON, outperforming traditional form parsers.
    \item We design a QA procedure that reduces field ambiguity through layout-aware and semantic-aware grouping, evaluated using two metrics: IR \emph{accuracy} and \emph{coverage}.
\end{itemize}

This project, including all hyperparameters and prompts used in our experiments and user interface, is publicly available on GitHub.\footnote{\url{https://github.com/dlacksthd94/Railway-Safety}}

\section{Related Works}

Key information extraction (KIE) aims to extract information in a structured format. The earliest approaches \cite{huang2022layoutlmv3pretrainingdocumentai, appalaraju2021docformerendtoendtransformerdocument, peng2022ernielayoutlayoutknowledgeenhanced, hong2022brospretrainedlanguagemodel} were layout-aware encoder models that jointly model textual tokens and 2D layout signals, such as Optical Character Recognition (OCR) results. However, because they were primarily trained on receipts and invoices, these models performed poorly on long, text-rich documents.
With the rise of generative models, end-to-end generative approaches \cite{kim2022ocrfreedocumentunderstandingtransformerdonut, cheng2022trie++endtoendinformationextraction, lee2023pix2structscreenshotparsingpretraining, tang2023unifyingvisiontextlayoutudop, yu2025omniparserv2structuredpointsofthoughtunified} emerged. They are context-aware, flexible in schema definition, and OCR-free. However, hallucination has emerged as a new challenge. And despite these advances, robust generalization to unstructured, text-rich government forms remains difficult.


\section{Proposed Pipeline}
We developed a pipeline that converts Form 57 into a JSON schema using VLMs with reduced hallucination and performs category-guided QA over news articles, as illustrated in Figure~\ref{fig: pipeline}. To evaluate the system, we constructed a dataset by scraping and preprocessing news articles and aligning them with official FRA records, which enabled measurement of \emph{accuracy} and \emph{coverage}.


\subsection{Key Information Extraction}

\begin{algorithm}[!ht]
\caption{Document Transcription and Entry Grouping}
\label{alg:KIE}
\KwIn{Form document $doc$, number of samples $N$}
\KwOut{Final JSON transcription $T_{\text{final}}$, final grouping $G_{\text{final}}$}

\For{$i \gets 1$ \KwTo $N$}{
  \Repeat{
    $T_i \gets \textsc{GenerateTranscription}(doc)$
  }{
    \textsc{ValidateTranscriptionFormat}($T_i$)
  }
}

$T_{\text{samples}} \gets \{T_1, \dots, T_N\}$\;

\Repeat{
  $T_{\text{final}} \gets \textsc{MergeTranscriptions}(doc, T_{\text{samples}})$
}{
  \textsc{ValidateTranscriptionFormat}($T_{\text{final}}$)
}

\For{$i \gets 1$ \KwTo $N$}{
  \Repeat{
    $G_i \gets \textsc{GenerateGroups}(doc, T_{\text{final}})$
  }{
    \textsc{ValidateGroupsFormat}($G_i$)
  }
}

$G_{\text{samples}} \gets \{G_1, \dots, G_N\}$\;

\Repeat{
  $G_{\text{final}} \gets \textsc{MergeGroups}(doc, T_{\text{final}}, G_{\text{samples}})$
}{
  \textsc{ValidateGroupsFormat}($G_{\text{final}}$)
}

\Return $(T_{\text{final}}, G_{\text{final}})$\;
\end{algorithm}

Given a form image, we prompt a VLM to transcribe the form in a predefined JSON schema.
We employ two main strategies to generate precise and thorough transcriptions:


\begin{enumerate}
    \item\label{s1} \textbf{Sample aggregation.} To mitigate hallucinated or omitted elements, we apply a sample-and-aggregate procedure: prompt the VLM to generate $N$ independent transcriptions of the entire form, each validated on whether it strictly follows the provided JSON schema (\ref{schema2}), and then synthesize them into a more complete final version using the VLM.
    
    \item\label{s2} \textbf{Human-centric schema.} We employ a JSON schema that requires the model to identify every subfield within each field, since each subfield corresponds to an actual area where a user writes or marks. This subfield-level approach (\ref{schema2}), rather than a naive field-level approach (\ref{schema1}), helps avoid partial extraction. For example, when transcribing the “6. Time of Accident/Incident” field, the naive approach often captures only the "AM/PM" checkbox while missing the "Time" textbox -- an empty space that is easy to overlook.
\end{enumerate}

However, many fields in Form 57 still remain ambiguous in isolation and exhibit interdependencies that can only be clarified in the context of surrounding fields. To address this, we ask the VLM to categorize fields into groups by jointly considering their visual layout and semantic meanings. Example groups include \emph{time \& location}, \emph{highway user}, \emph{train}, \emph{hazardous materials}, \emph{casualties}, and \emph{environment}.
Algorithm~\ref{alg:KIE} formalizes the overall process. 

\subsection{Information Retireval}

We perform QA for each group in $G_{\text{final}}$, using news articles as context. The LLM is instructed to produce answers according to each field’s type -- text, numeric, or single-choice. If a field isn't answerable from the news article, the system outputs \texttt{Unknown}.

\subsection{Evaluation Dataset Construction}

For each accident record in the FRA \href{https://data.transportation.gov/Railroads/Highway-Rail-Grade-Crossing-Incident-Data-Form-57-/7wn6-i5b9/about_data}{\emph{Highway–Rail Grade Crossing Incident Dataset}}, we searched for corresponding news reports by querying its location (state, county, city) and date range (0 to +7 days). We used several Python-based news content extractors -- \texttt{newspaper3k}, \texttt{trafilatura}, \texttt{readability}, and \texttt{goose3} -- under two strategies: direct HTTP requests and active crawling to handle client-side rendering and lazy-loaded content.

We initially scraped 1,707 articles from major news channels such as CBS and ABC, as well as from local media. However, nearly two-thirds were irrelevant to railway accidents -- covering other types of accidents, recalling railway tragedies from the distant past, or consisting primarily of boilerplate text. Due to a public concern of railroad crossing safety, we retained only articles reporting recent train–vehicle or train–pedestrian collisions.

To evaluate IR \emph{accuracy}, we linked news articles to official FRA records using temporal proximity (date/time window), spatial information (county, city, highway name), highway user details (sex and age), and casualty counts. The final dataset consists of 266 matched news–record pairs. From these, we sampled 50 pairs and manually annotated whether each field was answerable from the article, to enable measurement of \emph{coverage}.

\section{Evaluation}

We evaluated our two parsing strategies on \texttt{Qwen3-VL-32B-Instruct} \cite{bai2025qwen25vltechnicalreport}, \texttt{ChatGPT o4-mini}, and \texttt{Gemini 2.5 Flash} which can reliably parse the complex Form 57. Table~\ref{tab: transcription} reports the average number of KIE errors across four simulations. We define one  error as a field containing at least one incorrectly transcribed value. Among the 66 fields in Form 57, the average KIE error count was significantly lower, when both strategies were applied. AWS Textract, one of the best commercial parsers, produced lots of errors.

\begin{table}
    \caption{The impact of parsing strategies to KIE. Strategies 1 and 2 are referring to the sample aggregation and human-centric schema induction, respectively.}
    \label{tab: transcription}
    \begin{center}
        \begin{tabular}{cc|c}
            \toprule
            \textbf{KIE model} & \textbf{Parsing strategies } & \textbf{\# of errors (out of 66)} \\
            \midrule

            \texttt{AWS Textract} & - & 25 \\
            \cmidrule{1-2}

            \multirow{3}{*}{\texttt{o4-mini}} & \ref{s1} & 4.75 $\pm$ 1.78 \\
            & \ref{s2} & 2.75 $\pm$ 0.82 \\
            & \ref{s1} + \ref{s2} & 1.25 $\pm$ 0.43 \\

            \cmidrule{1-2}
            
            \multirow{2}{*}{\texttt{2.5 Flash}} & \ref{s2} & 1.25 $\pm$ 0.02 \\
            & \ref{s1} + \ref{s2} & \textbf{1} $\pm$ 0.01 \\
            
            \bottomrule
        \end{tabular}
    \end{center}
\end{table}

We measured performance using two application-centric metrics. \emph{Accuracy} measures per-field exact match between the QA output from the news articles and the ground-truth label from the linked FRA records. To avoid spurious mismatches, we adopt different judging rules: text-type answers are judged by an LLM to determine whether the response fuzzy-matches the ground-truth label. For certain digit-type fields such as time and speed, we allow a reasonable tolerance (e.g., within one hour or 10 MPH). \emph{Coverage} measures how much information the QA model attempts to retrieves. It is defined as $N_{\text{attempted}} / N_{\text{answerable}}$, where $N_{\text{answerable}}$ denotes the number of retrievable fields from a news article, and $N_{\text{attempted}}$ denotes the subset of those fields for which the QA model didn't respond with \texttt{Unknown}.

\begin{table}
    \caption{QA performance}
    \label{tab: perf}
    \begin{center}
        \begin{tabular}{ccc|cc}
            \toprule
            \multirow{2}{*}{\textbf{Pipeline}}
              & \multirow{2}{*}{\textbf{KIE model}}
              & \multirow{2}{*}{\textbf{QA batch}}
              & \multicolumn{2}{c}{\textbf{Metrics}} \\
            \cmidrule{4-5}
            & & & \textbf{Accuracy} & \textbf{Coverage} \\
            \midrule
            

            \textbf{Baseline 1} & - & Single & 0.52 & \underline{0.91} \\
            \cmidrule{1-3}
            
           
            \multirow{3}{*}{\textbf{Baseline 2}} & \texttt{Qwen3-VL} & \multirow{3}{*}{Single} & 0.55 $\pm$ 0.00 & 0.89 $\pm$ 0.00 \\

            & \texttt{o4-mini} &  & 0.56 $\pm$ 0.02 & 0.88 $\pm$ 0.03 \\

            & \texttt{2.5 Flash} &  & 0.57 $\pm$ 0.01 & 0.89 $\pm$ 0.01 \\
            \cmidrule{1-3}
            
           
            \multirow{3}{*}{\textbf{Baseline 3}} & \texttt{Qwen3-VL} & \multirow{3}{*}{All} & \textbf{0.71} $\pm$ 0.01 & 0.90 $\pm$ 0.00 \\

            & \texttt{o4-mini} &  & 0.67 $\pm$ 0.03 & 0.85 $\pm$ 0.06 \\

            & \texttt{2.5 Flash} &  & 0.70 $\pm$ 0.00 & 0.89 $\pm$ 0.01 \\
            \cmidrule{1-3}
            
           
            \multirow{3}{*}{Ours} & \texttt{Qwen3-VL} & \multirow{3}{*}{Group} & 0.67 $\pm$ 0.01 & 0.95 $\pm$ 0.00 \\

            & \texttt{o4-mini} &  & 0.65 $\pm$ 0.02 & 0.89 $\pm$ 0.08 \\

            & \texttt{2.5 Flash} &  & 0.67 $\pm$ 0.01 & \textbf{0.96} $\pm$ 0.01 \\
            \bottomrule
        \end{tabular}
    \end{center}
\end{table}

Using \texttt{Phi-4} as the QA model, we compare our approach against three alternatives: \textbf{Baseline 1.}  \emph{Field names from the FRA records → per-field single QA.} We construct the field list directly from the simplified column names and options in the FRA records, which is one of the simplest ways to build such a system, and query each field over the news articles; \textbf{Baseline 2.} \emph{VLM transcription → per-field single QA.} We use our VLM-derived transcription but ask questions one by one, without grouping; \textbf{Baseline 3.} \emph{VLM transcription → all-in-one QA.} This setting is identical to \textbf{Baseline 2} except that all fields are queried in a single prompt.


Table~\ref{tab: perf} reports overall accuracy and coverage. \textbf{Baseline 1} produces the worst result, likely because the simplified field names don't represent the original semantics of Form 57, despite having accurate choice lists and answer types. \textbf{Baseline 2} also underperforms our method, indicating that improved transcription alone is insufficient without semantic disambiguation.

A comparison between \textbf{Baseline 3} and our method highlights the importance of balancing prompt density. While both leverage the VLM-derived transcription and address field ambiguity, \textbf{Baseline 3} suggests that increased contextual complexity causes missing more answerable information. In contrast, our grouped QA strategy achieves higher coverage with a modest reduction in accuracy. Although this trade-off is evident, from an operational perspective it is critical that railway workers receive as much actionable information as possible during the early stages of an incident. In this context, our approach provides broader access to accident information.


\begin{table}
    \caption{QA accuracy for different answer types}
    \label{tab: ans type}
    \begin{center}
        \begin{tabular}{c|ccc}
            \toprule
            \multirow{2}{*}{\textbf{Pipeline}} & \multicolumn{3}{c}{\textbf{Answer type}} \\
            \cmidrule{2-4}
            & \textbf{Single choice}& \textbf{Digit}& \textbf{Free text} \\
            
            \midrule
            
            \textbf{Baseline 1} & 0.57 & 0.44 & 0.57 \\
            
            \textbf{Baseline 2} & 0.59 $\pm$ 0.02 & 0.52 $\pm$ 0.01 & 0.61 $\pm$ 0.01 \\

            \textbf{Baseline 3} & 0.71 $\pm$ 0.02 & 0.68 $\pm$ 0.01 & 0.75 $\pm$ 0.03 \\
            
            Ours & 0.69 $\pm$ 0.01 & 0.64 $\pm$ 0.03 & 0.64 $\pm$ 0.03 \\
            \bottomrule
        \end{tabular}
        \label{tab1}
    \end{center}
\end{table}

We further analyze IR accuracy by answer type in Table~\ref{tab: ans type} for \texttt{Gemini 2.5 Flash}. The relatively poor performance on digit fields is unsurprising, as numerical details in news articles (e.g., casualties or passenger counts) are often tentative and updated over time. The particularly poor performance of \textbf{Baseline 1, 2} on digit, compared to single-choice fields, can be explained by the fact that predefined option lists help disambiguate the meaning of a field, whereas digit field provide no such cues. This finding supports our assumption that ambiguity strongly affects IR.


Finally, we analyze the fields that are most frequently retrieved incorrectly. The fields with the highest error rates are "24. Type of Equipment Consist" and "17. Equipment", which achieve accuracies of 0.48 and 0.56, respectively, despite being single-choice fields. This degradation is largely attributable to the breadth and specificity of their choice sets: many of the fine-grained equipment categories and codes defined in Form 57 rarely appear explicitly in news articles, making correct mapping difficult even when partial evidence is present.


\section{Conclusion}
We presented a practical system for real-time population of FRA Form 57 from news articles, built around two core ideas: (i) VLM-driven form transcription and (ii) ambiguity mitigation through grouped QA. To support rigorous assessment, we constructed an evaluation dataset by scraping news reports and matching them to official FRA records. Evaluation using two application-centric metrics -- \emph{accuracy} and \emph{coverage} -- demonstrated that our pipeline substantially improves performance compared to approaches that omit these strategies.
Leveraging the reasoning capabilities of VLMs, this approach can be applied to other highly unstructured forms such as \href{https://railroads.dot.gov/sites/fra.dot.gov/files/2020-07/FRA\%20F\%206180.71.pdf}{Form 71} and \href{https://railroads.dot.gov/sites/fra.dot.gov/files/2019-09/FRA\%20F\%206180.55a.pdf}{Form 55a}. Future work includes incorporating additional data sources to address underreported incidents and extending the system beyond California to nationwide deployment.


\begin{acks}
We would like to thank the Riverside County Transportation Commission (RCTC), especially Sheldon Peterson and Paul Mim Mack, for constructive discussions. Research was supported by the National Science Foundation under grant no. 2431569 and by the University
Transportation Center for Railway Safety (UTCRS) at UTRGV through the USDOT
UTC Program under Grant No. 69A3552348340.
\end{acks}

\bibliographystyle{ACM-Reference-Format}
\balance
\bibliography{refs}

\appendix

\section{JSON schema}\label{schema}

\subsection{Schema not considering human perspective}\label{schema1}
\begin{verbatim}
"name": "<field name>",
"answer_type": "<text/digit/choice>",
"choices": {
    "<choice code>":  "<choice label>",
}
\end{verbatim}

\subsection{Schema considering human perspective}\label{schema2}
\begin{verbatim}
"name": "<field name>",
"answer_places": {
    "<answer place name in a few words>": {
        "answer_type": "<text/digit/choice>",
        "choices": {
            "<choice code>": "<choice name>",
        },
    },
}
\end{verbatim}


\end{document}